\documentclass[twocolumn]{jpsj2}
\usepackage{epsfig}
\def\gsim{\; $\raise0.3ex\hbox{$>$}\llap{\lower0.8ex\hbox{$\sim$}}$\;}
\def\lsim{\; $\raise0.3ex\hbox{$<$}\llap{\lower0.8ex\hbox{$\sim$}}$\;}

\title{On Calculation of Vector Spin Chirality for Zigzag Spin Chains}
\author{Kouichi Okunishi}
\inst{
Department of Physics, Faculty of Science, Niigata University, Igarashi 2, 950-2181, Japan.
}
\date{\today}

\abst{ We calculate the vector spin chirality for $S=1/2$ zigzag spin chains having U(1) symmetry, using the density matrix renormalization group combined with  unitary transformation.
We then demonstrate the occurrence of the chiral order  for the zigzag XY chain and  discuss the associated phase transition.
 The results are consistent with the analysis based on the bosonization and the long distance behaviour of the chirality correlation function.
For the $S=1/2$ zigzag Heisenberg chain in a magnetic field, we also verify the chiral order  that is predicted by the effective field theory and the chirality correlation function,  and then determine its magnetic phase diagram.
}

\kword{zigzag chains, vector chirality, spin current, spin rotation, DMRG}

\begin{document}
\maketitle

\section{introduction}

Recently the vector spin chirality in low dimensional quantum systems has been attracting much attention.
In particular, the $S=1/2$  zigzag chains have been providing considerably interesting physics associated with the ordering of vector spin chirality at the zero temperature\cite{chubukov,nersesyan,hikihara,lecheminant};
A strong quantum fluctuation breaks the classical helical order and the low-energy physics is described by disordered ground states such as the Tomonaga-Luttinger(TL) liquid state and dimerized state.
However, a ``hidden'' vector chiral order can be realized  in the antisymmetric sector, accompanying the spontaneous breaking of $Z_2$ symmetry originating from the double-chain nature of the zigzag lattice.
In Refs.[\cite{korezhuk,mccolloch,hikihara2}],  it was also shown that the vector chiral order is also induced for the zigzag Heisenberg spin chain by a uniform magnetic field.
An interesting point in the chiral order of the zigzag chains is that it can play an important role in the experimental situation of multiferroics through the spin-orbit coupling,\cite{furukawa} although the chirality itself may not be detected by  uniform magnetization measurement.

The vector spin chirality we discuss here is defined as 
\begin{equation}
\hat{\kappa}_{ij}\equiv[\vec{S}_i\times\vec{S}_j]_z, \label{chiral}
\end{equation}
on the bond of the $i$- and $j$-th sites, where  $\vec{S}$ is the $S=1/2$ spin operators and the spin quantization axis associated with the U(1)-rotational symmetry  is in the $z$-direction.\cite{ms,kawamura}
This quantity is also mentioned as a spin current on emphasizing the quantum nature of the U(1)-symmetric spin chain: $\hat{J}_{ij}\equiv J\hat{\kappa}_{ij}\equiv -iJ/2(S_i^+S_j^--S_i^-S_j^+)$, where $J$ is the exchange coupling.
A theoretical difficulty in investigating the chiral order, particularly in numerical approaches, is that, in the convention of the Pauli matrix,  all matrix elements of eq. (\ref{chiral}) are pure imaginary, while the wavefunction can be  represented by real numbers, since the spin chain Hamiltonian is a real symmetric matrix. 
This implies that an external field conjugate to the chirality is exactly zero even at the level of numerics, and then the expectation value of eq.  (\ref{chiral}) must be zero even in the occurrence of a spontaneous order of the chirality.
In the previous numerical analysis, accordingly, the properties of the spin chirality were extracted from the long-distance behavior of the chirality-chirality correlation function.\cite{hikihara,mccolloch,hikihara2,yoshi}
However, the groundstate expectation value of the chirality operator itself is clearly desired for the precise analysis of the chiral order.

In analogy with the Josephson current in superconductivity, a phase of spin, which is associated with a spin rotation angle,  can be expected to be crucial for spin current.
This suggests that a certain spin rotation makes the  spin chirality more clearly observable and enables  us to directly calculate the spin chirality.
In this paper, we actually introduce a unitary transformation that makes the $Z_2$ symmetry associated with the spin chirality more visibly, and then calculate the spin chirality order parameter, using the density the matrix renormalization group (DMRG)\cite{dmrg}.  
On this basis, we discuss the various properties associated with the chiral order for zigzag spin chains.

\begin{figure}[tb]
\begin{center}
\epsfig{file=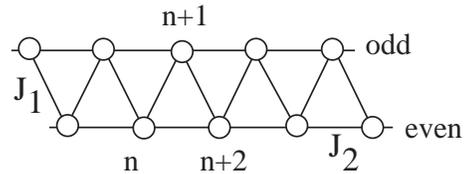,width=6cm}
\caption{Zigzag lattice; $n$=even and odd sites form the double chain structure.}
\end{center}
\end{figure}

The $S=1/2$ zigzag chain is one of the most essential 1D quantum spin systems having the frustration effect; its Hamiltonian is given by
\begin{eqnarray}
{\cal H}= {\cal H}_{1} + {\cal H}_{2} \label{hamiltonian}
\end{eqnarray}
with
\begin{eqnarray}
{\cal H}_{1}&= &J_1 \sum_{n} [S_{n}^xS_{n+1}^x + S_{n}^yS_{n+1}^y +\Delta S_{n}^zS_{n+1}^z ],  \\
{\cal H}_{2} &=&J_2 \sum_{n} [S_{n}^xS_{n+2}^x + S_{n}^yS_{n+2}^y +\Delta S_{n}^zS_{n+2}^z ],
\label{zigzag}
\end{eqnarray}
where $J_1$ and $J_2$ denote the nearest and next-nearest neighbor couplings, respectively, and $\Delta$ is the anisotropy parameter. 
As in Fig. 1, the $n=$even and odd sites form the double-chain structure without losing the translational invariance.
In this paper, we basically consider $J_2 > J_1 > 0 $ and thus introduce the notation $\alpha=J_1/J_2$ for simplicity.
The symmetry of the zigzag spin chain and vector chiral order was analyzed in Ref.[\cite{lecheminant}].
Here, we would like to mention that the $Z_2$ symmetry corresponding to the parity between $n$=even and odd chains is important for the vector chiral order.

This paper is organized as follows. 
In \S 2, we introduce the unitary transformation corresponding to the $\pi/2$ rotation of the spin phase, which makes the $Z_2$ symmetry of the system more transparent.
As a similar but different approach, we also introduce phase modulation into the system, which can be regarded as a field conjugate to the chirality. 
In \S 3, we calculate the expectation value of the  chirality operator for the zigzag XY chain ($\Delta=0$) in the bulk limit, using the infinite system size method of DMRG accelerated by the recursion relation for the wavefunction.\cite{pwfrg} 
We then find that the chirality order parameters basically agree with the previous results based on the bosonization and the analysis of the correlation functions.
We also discuss the chiral order in the context of phase modulation in \S 2.
In \S 4, we calculate the spatial distribution of the local chirality for the zigzag Heisenberg chain in a magnetic field. 
We then show that the chiral order is actually induced for a certain region of the magnetic field and $\alpha$, which is also consistent with the previous weak coupling theory\cite{korezhuk} and the correlation function analysis.\cite{mccolloch}
We then find that the transition between the chiral-ordered phase and the two-component TL liquid\cite{cusp} may still appear, and finally determine the magnetic phase diagram in addition to the phase boundary obtained  previously\cite{zigzag,zigzag2}~.
In \S5,  we summarize the results.

\section{chirality and spin phase}

As was described in the introduction, the main difficulty in the numerical calculation of the chirality order parameter is in that the Hamiltonian (\ref{hamiltonian}) is a real symmetric matrix, which is not suitable for dealing with the phase of spins;
the spin current is caused by the phase difference between $n=$even and odd sites,  which is associated with the $Z_2$ symmetry breaking of the sign (direction) of the current.
We thus try a spin rotation to control the relative phase of the spins on even and odd sites.

\subsection{unitary transformation}
Let us write the spin rotation operator around the $z$-axis at the site $n$ as 
$R^z_n(\theta)\equiv e^{i\theta S_n^z}$,
so that $R_n^z(\theta) : S_n^+ \to e^{i\theta} S_n^+$ and  $S_n^z \to S_n^z$.
Then we define a unitary operator as,
\begin{equation}
U^z(\frac{\pi}{2}) = \prod_{n \in \rm even} R_n^z(\frac{\pi}{2}),\label{unir}
\end{equation}
which rotates spins at $n=$even sites by $\pi/2$ around the $z$ axis.
Then ${\cal H}_1$ is transformed into
\begin{eqnarray}
\tilde{\cal H}_1 = \sum_n \frac{(-)^n}{2i}(S^+_n {S}^-_{n+1} -S^-_n {S}^+_{n+1}) +\Delta S^z_n{S}^z_{n+1},\label{mappedh1}
\end{eqnarray}
while ${\cal H}_2$ is unchanged.
Note that, if $\theta=-\pi/2$ rotation is employed instead of eq. (\ref{unir}), the sign at the XY term in $\tilde{\cal{H}}_1$  is inverted.
Thus, the sign of the spin rotation is closely related to the $Z_2$ symmetry of the spin chirality.

The original Hamiltonian (\ref{hamiltonian}) is invariant under the one-site translation operation: $n \to n+1$, but a naive translation of the transformed Hamiltonian (\ref{mappedh1}) changes the sign in front of the XY term, which implies that the explicit translational invariance is hidden by the unitary transformation.
However, the translation combined with an exchange  of the label of even and odd sites preserves eq. (\ref{mappedh1}) including the sign of the XY term.
In the representation of eq. (\ref{mappedh1}),  the other symmetry operations presented in Ref. [\cite{lecheminant}] are also preserved.
Then an essential point is that  the $Z_2$ symmetry in the Hamiltonian becomes more explicit, which is expected to be suitable for  numerical treatment.

According to (\ref{unir}), the chirality operator is also transformed into
\begin{equation}
\tilde{\kappa}_{n,n+1}= \frac{(-)^n}{2}(S^+_n {S}^-_{n+1} +S^-_n {S}^+_{n+1}).
\label{mappedk}
\end{equation}
As can be seen for eq. (\ref{mappedh1}), the local chirality operator (\ref{mappedk}) is invariant for the one-site translation combined with the exchange of the label of even and odd sites. 
Then an important point is that the sign of eq. (\ref{mappedk}) is assigned staggeredly, so that the sign of the uniform operator $S^+_n {S}^-_{n+1} +S^-_n {S}^+_{n+1}$  should be  broken staggeredly to produce a uniform chiral order. 
In addition to the above, we just note that the form of the chirality operator and the XY term in ${\cal H}_1$ exchanges with each other in this unitary transformation, which may interestingly suggest a kind of duality for the spin chirality.

\subsection{phase modulation}

We consider the Hamiltonian (\ref{hamiltonian}) again by introducing a phase modulation for the XY term. 
Such an extra phase factor is often discussed in the context of the persistent current associated with the gauge flux for the periodic ring.
Here, we assign the modulation only for the nearest-neighbor bonds, such that
\begin{eqnarray}
{\cal H}_{1}(\phi)& \equiv& J_1\sum_n \big[ \frac{1}{2}(e^{i\phi}S^+_n S^-_{n+1} + e^{-i\phi}S^-_n S^+_{n+1})\nonumber \\
&& \left.  +\Delta S^z_nS^z_{n+1}\right],
\label{fluxh1}
\end{eqnarray}
where $\phi$ is a phase variable.
Then the operator of the vector chirality order parameter can be obtained as
\begin{equation}
{\cal O}_\kappa \equiv \frac{1}{L} \sum_n\hat{\kappa}_{n,n+1}=  -\frac{1}{J_1 L}\left.\frac{\partial{\cal H}(\phi)}{\partial \phi}  \right|_{\phi=0},
\end{equation}
where $L$ is the length of the chain.
Writing the ground state energy per bond with the modulation $\phi$ as $E(\phi)$, we can then calculate the chirality order parameter
\begin{equation}
\langle {\cal O}_\kappa\rangle = -\left. \frac{1}{J_1}\frac{\partial E(\phi)}{\partial \phi}  \right|_{\phi=0},
\end{equation}
indicating that $\phi$ can be regarded as the field conjugate to the chirality order parameter.
We can thus discuss  $E(\phi)$ and $\phi$, similarly to the relation between the free energy and the magnetic field for the Ising model;
if the spontaneous chiral order that breaks the $Z_2$ symmetry occurs, it can be extracted as a singularity of $E(\phi)$ in the limit $\phi\to \pm 0$.

Here, we want to comment on the implication of the phase modulation to the experiment;
the imaginary part of the phase factor in eq. (\ref{fluxh1}) can be casted as the Dzyaloshinskii-Moriya (DM) interaction\cite{dm}, by renormalizing the coupling $J\cos\phi \to J$ and $J\sin\phi\to D_z$, where $D_z$ is the $z$ component of the DM vector.
In this sense, it is possible to control the chiral order through the DM interaction in the experimental situation.

\subsection{DMRG}

In order to investigate the transformed Hamiltonian, we use DMRG.\cite{dmrg}
Although numerical variables become complex numbers, the Hamiltonian is of course an Hermite matrix, so that we  have no essential difficulty in DMRG computation.
For the bulk properties of the chiral order, we particularly employ the product wavefunction renormalization group\cite{pwfrg}, which is an infinite-size algorithm of DMRG assisted by the recursion relation for the wavefunction.
After  sufficient iterations, we calculate the expectation value of the  chirality operator in the bulk limit as a solution of the self-consistent equation of the DMRG recursion relation;
since the $Z_2$ symmetry is directly represented as a sign in eq. (\ref{mappedh1}),  one of the broken symmetry states can be selected in the iteration of DMRG, owing to the perturbation from the numerical round-off error, or an initial condition of the DMRG.
We also perform the finite system size calculation of DMRG for a finite but long chain, for which the broken symmetry state is eventually achieved owing to the numerical perturbation as well.
We then discuss the spatial distribution of the chirality.

Here we should mention about the eigenvalue of the density matrix in DMRG.
For the chiral-ordered state of the original Hamiltonian (\ref{zigzag}), the density-matrix spectrum has a trivial two-hold degeneracy,  where the superposition of the two equivalent state originating from the $Z_2$ symmetry is observed. 
For the transformed Hamiltonian (\ref{mappedh1}), however, we have found that such  two-hold degeneracy disappears in the following DMRG computation.
This implies that one of the broken symmetry states can be extracted properly.

\section{zigzag XY chain}

For the zigzag XY chain, the chiral-ordered state was analyzed by the bosonization\cite{nersesyan} and the ground state phase diagram for the chiral order was obtained  in Ref. [\cite{hikihara}], on the basis of  the long-distance dumping of the chirality correlation function.
The phase boundary is $\alpha_c \simeq 0.794$; the chiral order with the gapless magnetic excitation appears for $\alpha <0.794$, while the dimer gapped state without the chiral order is located in $\alpha >0.794$.\cite{dimer}

\subsection{chirality order parameter}

\begin{figure}[bt]
\begin{center}
\epsfig{file=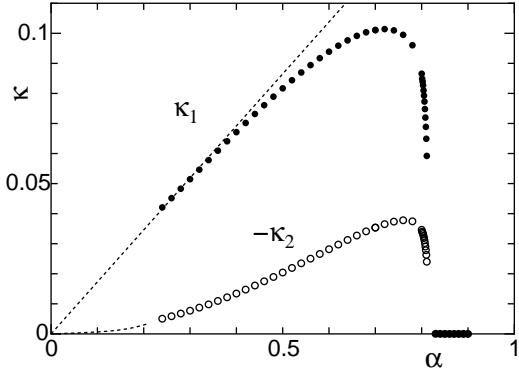,width=7cm}
\caption{Vector chirality for the XY model. Solid and open circles indicate $\kappa_{1}$ and $\kappa_{2}$. The broken line indicates $\kappa=1.73 \alpha$, which is a guide for the eyes.\label{fig2}
}
\end{center}
\end{figure}

Since we use the infinite DMRG, the bulk values of the chirality is measured as an expectation value of the local chirality operator at the center of the chain.
In the following, we thus denote the bulk expectation values of the chiralities for the nearest-neighboring spins and next-nearest-neighboring spins as  
\begin{equation}
\kappa_1 \equiv \langle \hat{\kappa}_{n,n+1}\rangle\quad {\rm and}\quad  \kappa_2\equiv \langle \hat{\kappa}_{n,n+2}\rangle, 
\end{equation}
respectively. 
Figure \ref{fig2} shows the results for $\kappa_1$ and $\kappa_2$.  
We can obtain sufficiently reliable results within the number of retained bases $m=600$ and the number of iterations $1200$, except in the very vicinity of $\alpha_c$ and in the region $\alpha< 0.24$.
The value of  $\kappa_{1}$ can be checked to be consistent with that obtained from the long-distance behavior of the chirality correlation function.
Note that much more bases is required for $\alpha< 0.24$, since DMRG is not efficient in the nearly decoupling limit of double chains.

In the figure, we can see that the chirality order parameter $\kappa_{1}$ develops almost linearly with respect to $\alpha$ for $\alpha <0.5$, where we also plot $\kappa = 1.73\alpha$ for  comparison.
The scale of the amplitude of the chiral order is basically proportional to the zigzag coupling $J_1$.
As $\alpha$ increases further,  the amplitude of  $\kappa_{1}$ decreases rapidly toward $\alpha\simeq 0.81$ after taking its maximum at around $\alpha\simeq 0.72$.
Detailed analysis yields this offset of the chiral order as $\alpha=0.81$(see the next paragraph), which is basically consistent with $\alpha_c=0.794$ on the basis of the correlation function.\cite{hikihara}
In addition, we can see that the chirality for the next nearest sites  $ \kappa_{2}$ appears in the same range of $\alpha$ as  $\kappa_{1}$;
the weak coupling analysis suggests the conservation of the global spin current\cite{current}
\begin{equation}
 J_1 \kappa_{1} = - 2J_2 \kappa_{2}. \label{cconserved}
\end{equation}
We have confirmed this relation in the DMRG result, indicating that the spin current is circulating in the unit triangles and thus there is no net spin current.
An interesting point in Fig. \ref{fig2} is that $\kappa_{2}$ has a curvature for $\alpha < 0.5$, in contrast to the linear behavior of  $ \kappa_{1} \sim 0.173 \alpha$.
According to eq. (\ref{cconserved}), this behavior is consistent with $\kappa_{2} = - 2\alpha J_2 \kappa_{1} \sim - \alpha^2$.

\begin{figure}[b]
\begin{center}
\epsfig{file=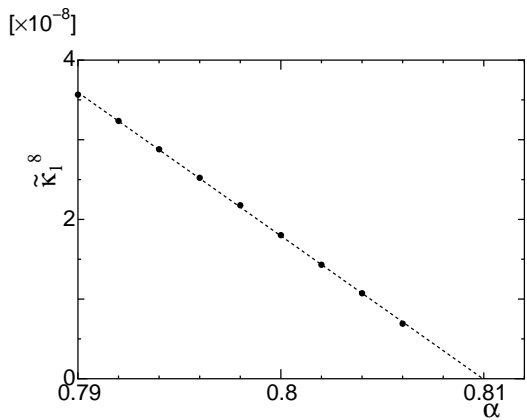,width=7cm}
\caption{${\tilde{\kappa}_1}^8$ vs. $\alpha$ plot for the zigzag XY model.
The broken line indicates the fitting result: ${\tilde{\kappa}_1}^8 =1.80\times 10^{-6}(0.810-\alpha)$
\label{fig3}
}
\end{center}
\end{figure}

Next, let us consider the critical behavior of the chiral order; 
for this purpose, we discuss the normalized chirality $\tilde{\kappa}_1\equiv \kappa_{1}/\alpha  $, because the energy scale of the chiral order is basically proportional to $\alpha$.
Since the $Z_2$ symmetry is broken, the expected universality of the transition is of  2D Ising class, for which the exponent of the order parameter is given by $\beta=1/8$.
We thus plot ${\tilde{\kappa}_1}^8$ against $\alpha$ in Fig. \ref{fig3}.
Then, we can verify the linear behavior of $\tilde{\kappa_1}^8$, indicating that the transition is of 2D Ising universality, as expected.
Moreover, a straightforward extrapolation yields  an estimation of the critical point, $\alpha_c = 0.81$.
This value is basically consistent with  $\alpha_c \simeq 0.794$ in Ref.[\cite{hikihara}], but a slight discrepancy can be seen.
Here, we should however recall that the double transition ---the present order-disorder transition of the chirality and the dimer-spin liquid transition--- is expected to occur at $\alpha_c$ within numerical accuracy.\cite{hikihara};
Thus, improvement in the accuracy in the vicinity of $\alpha_c$ is quite difficult in the previous approach of the correlation function.
We thus think that the results of the two approaches are consistent with each other. 
In addition, it should be noted that, if there is a coupling between these two modes, the critical phenomenon itself may be modified from the Ising universality.

\begin{figure}[t]
\begin{center}
\epsfig{file=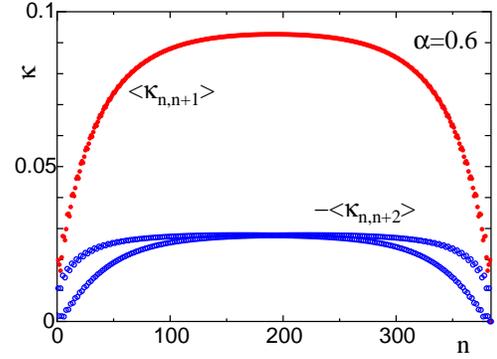,width=6.5cm}
\caption{(color online) Spatial distribution of chiralities for the zigzag XY model of $\alpha=0.6$. The system size is 384 sites. The solid and open circles indicate $\langle\hat{\kappa}_{n,n+1}\rangle$ and $\langle \hat{\kappa}_{n,n+2}\rangle$ respectively. \label{fig4} 
}
\end{center}
\end{figure}

In Fig. \ref{fig4}, we show the spatial distribution of the chirality on a finite-length chain of 384 spins, which is calculated by the finite-size DMRG up to $m=250$.
The  chirality is suppressed at the edges of the chain, but develops toward the center.
In the flat region around the center of the chain ($n\simeq 192$), we have $\langle \hat{\kappa}_{n,n+1}\rangle \simeq 0.0927$ and $\langle \hat{\kappa}_{n,n+2}\rangle\simeq 0.0278 $, which are consistent with the bulk values in Fig. \ref{fig2}.
In addition, we can see that the boundary effect more significantly appears for $\langle \hat{\kappa}_{n,n+2}\rangle $, which splits into  two branches near the edges depending on even and odd sites.
However, it can be confirmed that the conservation of the local spin currents
\begin{equation}
J_1\langle \hat{\kappa}_{n-1,n}\rangle - J_2 \langle \hat{\kappa}_{n-2,n}\rangle =J_1\langle \hat{\kappa}_{n-1,n}\rangle - J_2\langle \hat{\kappa}_{n,n+2}\rangle
\end{equation}
is satisfied at every site in the chain.

\subsection{phase modulation}

We next consider the effect of the phase modulation in eq. (\ref{fluxh1}).
In particular, we focus on the $\phi$-dependence of the groundstate energy and the vector chirality around $\phi=0$.
Using the infinite system size DMRG, we compute the gourdstate energy per bond $E(\phi)$ and the chirality order parameter $\kappa_1(\phi)$ in the bulk limit.
In practical DMRG computation, we use matrices obtained for a previous $\phi$ as an initial condition for $\phi$ shifted slightly.
Thus, we have a kind of ``hysteresis'' in the DMRG calculation, which enables us to deal with a metastable solution in the vicinity of $\phi=0$. 

\begin{figure}[tb]
\begin{center}
\epsfig{file=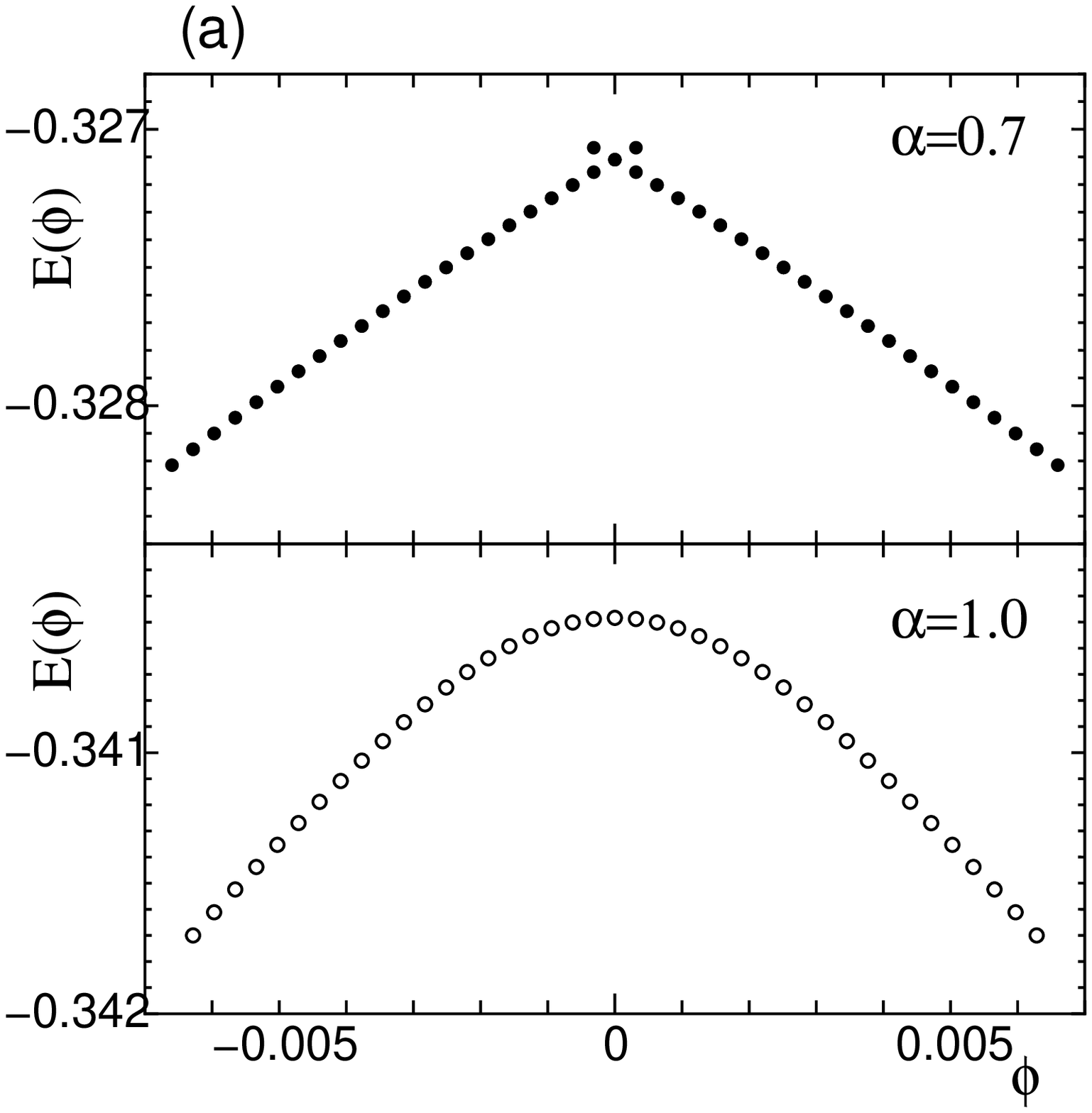,width=6cm}
\epsfig{file=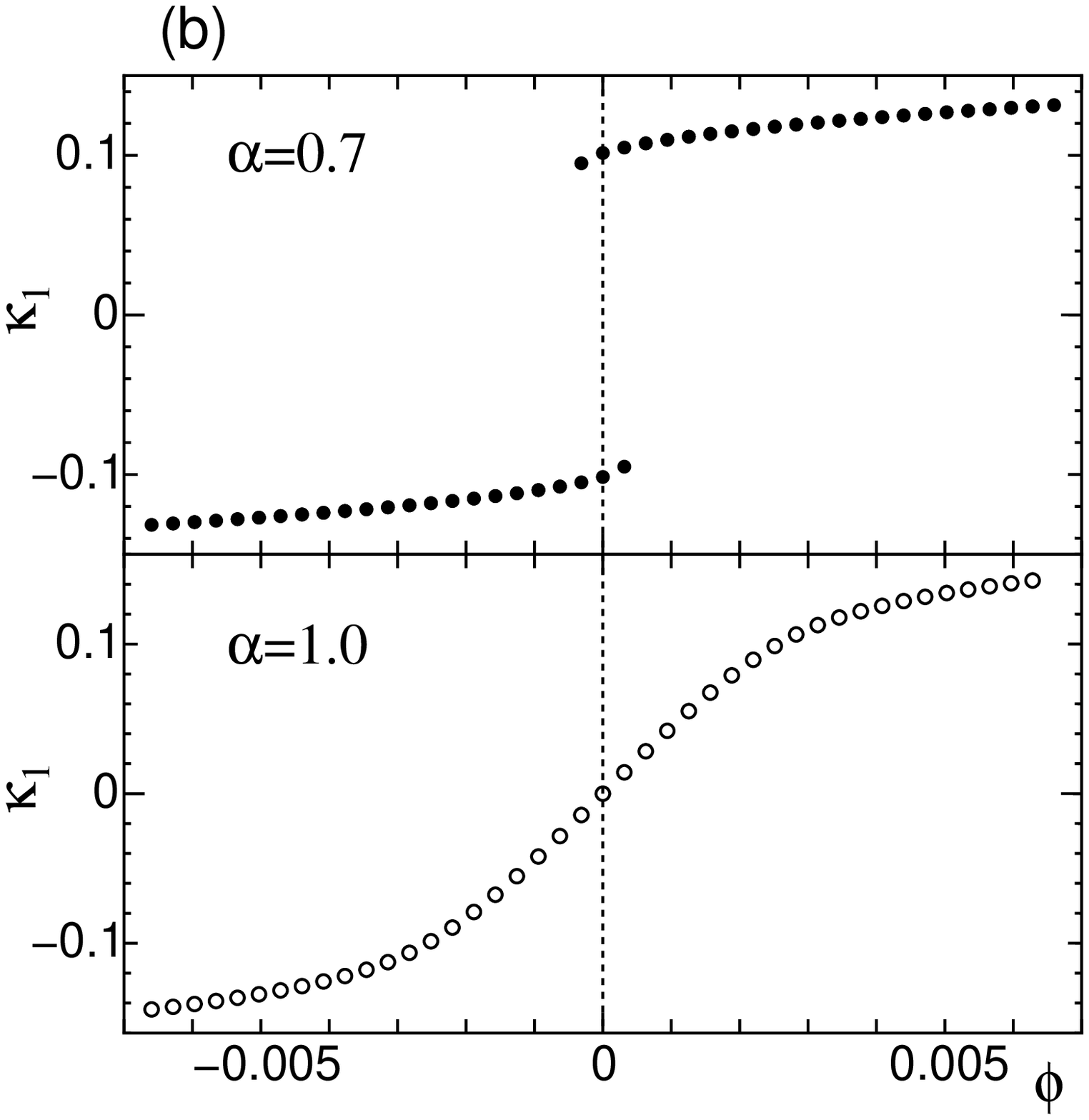,width=6cm}
\caption{(a) $E(\phi)$ for $\alpha=0.7$(solid circles) and $1.0$ (open circles). (b) $\kappa_{1}(\phi)$ for $\alpha=0.7$ (solid circles) and $1.0$ (open circles). The vertical broken line indicates $\phi=0$. \label{fig5}
}
\end{center}
\end{figure}

Figure \ref{fig5}(a) shows the groundstate energy per bond $E(\phi)$ and Fig. \ref{fig5}(b) shows the chirality $\kappa_{1}(\phi)$ for $\alpha=0.7$ and 1.0.
For $\alpha=0.7$, the energy has a cusp singularity at $\phi=0$, while for $\alpha=1.0$ the curve is smooth at $\phi=0$.
Accordingly, $\kappa_1$ for $\alpha=0.7$ has a discontinuity at $\phi=0$, while that for $\alpha=1.0$ is a continuous function of $\phi$.
The overhang branch at around $\phi=0$ for $\alpha=0.7$ indicates the metastable states associated with the first-order transition with respect to $\phi$.
Note that $\kappa_{1}$ at $\phi=0$ is agree with that obtained for the unitary-transformed Hamiltonian (\ref{mappedh1}) in the previous subsection.
It should also be  noted that the numerical derivative of $E(\phi)$ in Fig. \ref{fig5}(a) is consistent with $J_1\kappa_{1}$ in Fig. \ref{fig5}(b).
This implies that the chiral order is realized  for $\alpha<\alpha_c$ and that $\phi$ plays a role of the $Z_2$ symmetry breaking external field conjugate to the chirality.

\section{zigzag Heisenberg chain in magnetic field}

The magnetic phase diagram of the zigzag Heisenberg chain was obtained in Ref.[\cite{zigzag}].
However, the chiral order is not detected by the uniform magnetization measurement.
The bosonization study of the zigzag chain  reveals that the chiral order can emerge in the antisymmetric sector of double chains,  rather than the two-component TL liquid.\cite{korezhuk}
Recently, the correlation function analysis  presents that the chiral order actually exists in certain ranges of  magnetization \cite{mccolloch,hikihara2}.
Moreover, the phase boundary between the two-component TL liquid phase and the chiral phase can be suggested in the two-component TL liquid region of the previous phase diagram, although the magnetization curve has no  clear singularity.
In this section, we would like to address this problem on the basis of the direct calculation of the chirality order parameter based on eq. (\ref{mappedh1}). 

In order to deal with the chiral order in a magnetic field, we employ the finite system method of DMRG, since  the infinite system algorithm is not suitable for dealing with a magnetization-fixed system.
We calculate the spatial distribution of the local chirality order parameters rather than the chirality correlation function.

\begin{figure}[htb]
\begin{center}
\epsfig{file=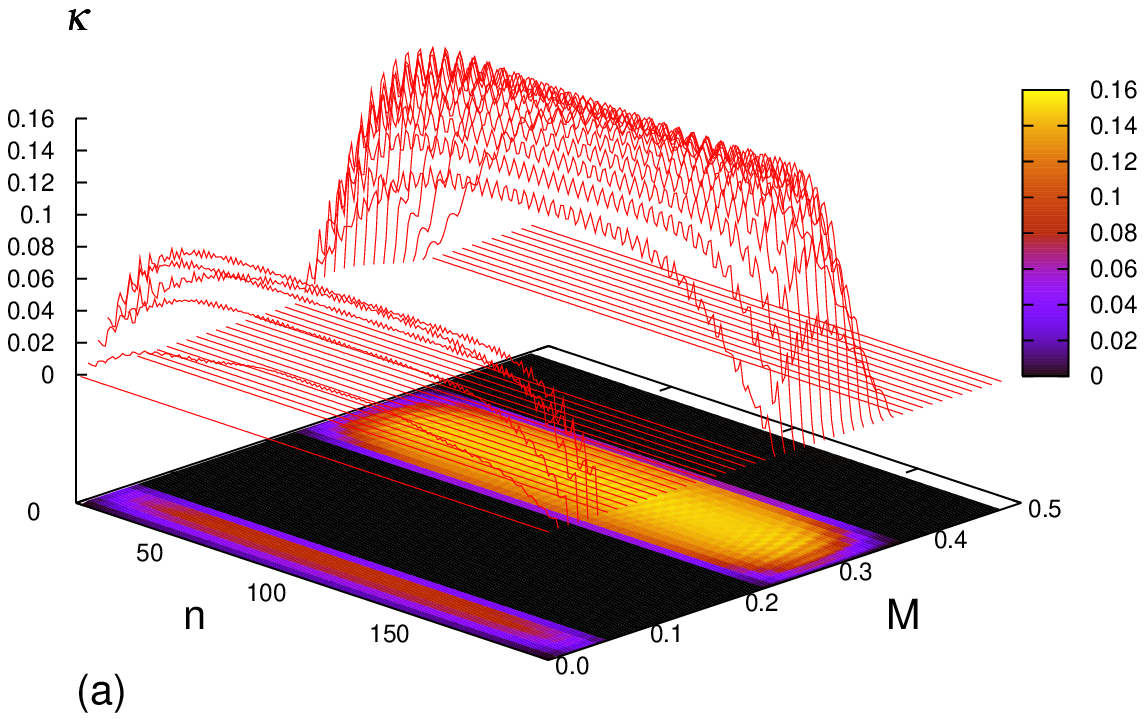,width=8.5cm}
\epsfig{file=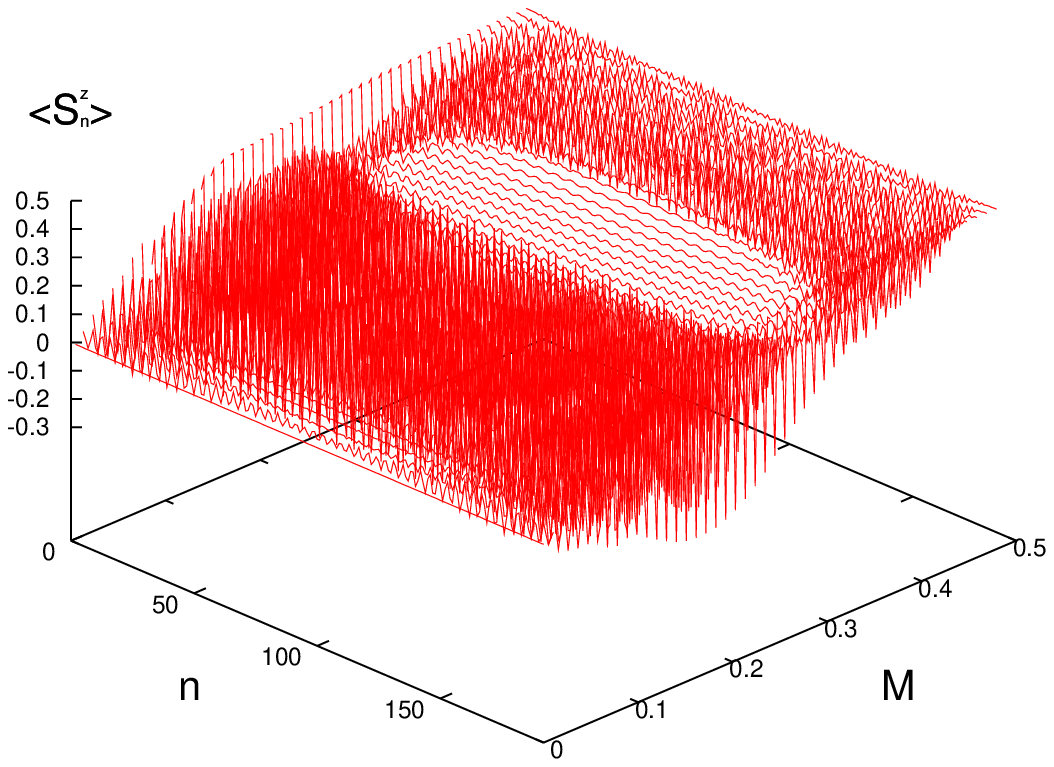,width=8.5cm}
\caption{(color online)  (a) Distribution of the amplitude of the vector spin chirality  $\langle\hat{\kappa}_{n,n+1}\rangle$ and (b) distribution of the local moment $\langle S_n^z\rangle$.  The horizontal axis indicates the  magnetization of the system $M$, and the site index $n$. The vertical axis represents $\kappa$ in (a) and $\langle S_n^z\rangle $ in (b). 
In (a), the line in the upper part indicates $\langle\hat{\kappa}_{n,n+1}\rangle$ for each magnetization, and the density plot at the bottom represents the projection of the amplitudes on the $n$-$M$ plane.
\label{fig6}
}
\end{center}
\end{figure}
 
\subsection{distribution of chirality}

In Fig. \ref{fig6}(a), we illustrate the spatial distribution of the local chirality order parameter as a function of the site index $n$ and the magnetization $M$, for a 192-site system of $\alpha = 1.0$. 
A line in the upper layer in Fig. \ref{fig6}(a) indicates  $\langle\hat{\kappa}_{n,n+1}\rangle$ for each magnetization, and the density plot at the bottom represents the projection of the amplitudes on the $n$-$M$ plane.
We can see that the chiral order appears in the low-magnetization region($0<M<0.05$) and in the upper-middle magnetization region ($0.24<M<0.365$)\cite{onset};
although the chiral order is suppressed at the edges, it develops toward the center of the chain.
Note that the amplitude of the chiral order for $0.24<M<0.365$ is larger than that in $0<M<0.05$.

\begin{figure}[htb]
\begin{center}
\epsfig{file=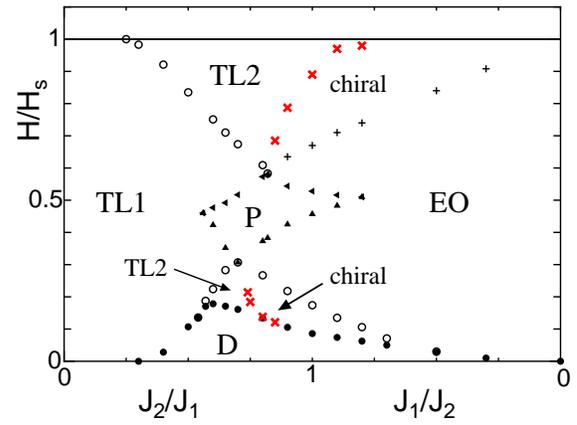,width=7.5cm}
\caption{(color online) Phase diagram of the zigzag Heisenberg chain in a magnetic field.
The chiral-ordered phase is added to the previous one in Ref.[\cite{zigzag}].
P: 1/3 plateau, 
D: dimer-ordered phase of zero magnetization, 
EO: even-odd phase where the oscillating behavior appears in the magnetization curve of finite size, depending on the total $S^z=$even or odd.
TL1: single component TL liquid, 
TL2: two-component TL liquid, and 
chiral: chiral-ordered phase. 
`` {\tt x} '' symbols denote the boundaries between the chiral-ordered phases and two-component TL phases.
\label{fig7}
}
\end{center}
\end{figure}

In comparison with the previous phase diagram\cite{zigzag}(see also Fig. \ref{fig7}), the region $0<M<0.05$ corresponds to the field range between the critical field $H_c$ and  the low field cusp of the magnetization curve,  where $H_c$ is the field that collapses the dimer gap.
The region $0.05 <M <0.24 $ corresponds to  the even-odd (EO) phase around the 1/3 plateau, which can be explained by  the non-chiral single component TL liquid attributed to the bound state of the domain wall excitations\cite{zigzag2}.
In the region $ M >0.24 $, the chiral order phase clearly appears, although the previous phase diagram would have indicated a two-component TL liquid.
Above $M\simeq 0.365$, however, the chiral order disappears again, in spite of the lack of explicit anomaly in the magnetization curve. 
This result is consistent with the correlation function analysis in Ref. [\cite{mccolloch}], suggesting that there is a new phase boundary of the chiral-ordered phase  at $M\simeq 0.365$.

In Fig. \ref{fig6}(b), we show the distribution of the local spin moment $\langle S^z_n\rangle $.
An important point is that the finite-size oscillation of  $\langle S^z_n\rangle $ is very suppressed in the chiral-ordered region,  and a small but clear $2k_F$ oscillation of the local spin moment $\langle S_n^z\rangle$ can be seen, where $2k_F$ is the Fermi wave number depending on the magnetization of the system.  
Since the bosonic field representation of  $\hat{\kappa}_{n,n+1}$  contains a $2k_F$ oscillating term in higher-order contributions, 
the $2k_F$ oscillation appears in  the chirality-chirality correlation function\cite{mccolloch}.
The present results  verifies that the same $2k_F$ oscillation can be seen in $\langle  \hat{\kappa}_{n,n+1}\rangle $ and  $\langle S_{n}^z\rangle$.
On the other hand, for $M>3.65$, the oscillation of $\langle S_n^z\rangle$ exhibits a complicated behavior;  
this suggests that the two-component TL liquid phase may be realized in $0.365 <M<1/2$, which can be explained by the double-well shape of the spin wave dispersion from the saturated state.\cite{cusp}

At the boundary between the chiral order phase and the two-component TL liquid phase, there is no clear anomaly in the magnetization curve.
However, the amplitude of the chirality order parameter  seems to change very steeply from a finite value to zero, which suggests that the transition might be of first order.
Of course, the present calculation is for a finite system size, where $M$ takes only discrete values.
Thus,  further research is clearly needed for the determination of the transition between the chiral phase and the two-component TL liquid.

\subsection{phase diagram}

We have performed DMRG calculations for 192-site systems of various $\alpha$ and $M$, and then summarize the results in Fig. \ref{fig7}: 
the magnetic phase diagram of the zigzag Heisenberg chain, by adding the present results for the chiral order into the previous magnetic phase diagram in Ref. [\cite{zigzag}].
The chiral order phase appears below the low-field cusp for $J_2/J_1 > 0.7$, and above the EO phase for $J_2/J_1 > 0.83$.
The boundaries of the chiral-ordered phases and the two-component TL phases are
denoted by `` {\tt x} ''.
The boundary of the upper chiral phase  approaches $J_2/J_1 \simeq 0.83$, at which the magnetization cusps merge into the 1/3 plateau.
As $J_2/J_1$ increases,  this boundary rises rapidly toward the saturation field.
Also below the 1/3 plateau,  the boundary of the lower chiral phase and the low-field cusp meets the 1/3 plateau at $J_2/J_1=0.7$.
As $J_2/J_1$ increases, then  this boundary approaches  the lower critical field corresponding to the dimer gap.
An interesting point is that the low-energy excitation  around the 1/3 plateau can be described well by the domain wall excitations carrying $S^z=\pm 1/3$ and their bound states.\cite{zigzag2}
Then $J_2/J_1 \simeq 0.7$ or  $0.83$ are located at points where the nature of the dominant excitations switches  between the  single domain wall and the domain wall bound state.
This suggests that the occurrence of the chiral order can be understood in the context of the nature of domain wall particles, which may provide another viewpoint on the mechanism of chiral ordering.

\section{summary and discussion}

We have discussed the vector chiral order for zigzag spin chains.
We have introduced a unitary transformation that rotates  spins on even sites by angle $\pm \pi/2$ around the $z$-axis.
This transformation makes the spin current nature of the phase difference more explicit and enables the direct numerical calculation of the chirality.
We also discuss the phase modulation as a conjugate field to the chirality.

For the $S=1/2$ zigzag XY chain, we have actually calculated the chirality order parameter in the bulk limit, using infinite-size DMRG.
The result has demonstrated that the chiral order emerges for $\alpha<0.81$, which is consistent with the known results.
However, we note that the unified analysis not only for the transitions of the chiral ordering but also for that of the dimer order is still an important issue.
For the $S=1/2$ zigzag Heisenberg chain in a magnetic field, we have also calculated the chirality profile of 192-site systems.
We then find that the chiral order is certainly realized above or below the cusp singularities of the magnetization curve, which quantitatively agree with  the correlation function analysis\cite{mccolloch,hikihara2}.
The chirality result also suggests that the transition between the chiral-ordered phase and the two-component TL liquid occurs, although there is no clear singularity in the magnetization curve.
We have finally illustrated the magnetic phase diagram of the zigzag Heisenberg chain  in Fig.\ref{fig7}.
The transition between the chiral phase and the two-component TL liquid seems to be of the first order within the 192-site result, accompanying a finite jump of the chirality. However, further researches of  the critical phenomena associated with the chiral order are clearly needed.

In order to discuss the physical background of the phase diagram, it may be important to see that the cusps and transition lines of the chiral order merge into the 1/3 plateau at $\alpha \simeq 0.7$ or $0.82$.  
The continuous field theory states that the chiral order is induced by the $Z_2$ symmetry breaking in the antisymmetric sector of two bosons originating from the double-chain structure of the zigzag lattice.
On the other hand, the domain wall excitation carrying $S^z=\pm 1/3$  also describes the single chain-double chain crossover around the 1/3 plateau\cite{zigzag2};
The nature of the low-energy excitations changes between the single domain wall and  the bound state of the domain walls at $\alpha \simeq 0.7$ or $0.82$.
Since the cusps and transition lines of the chiral order meet at the same point, the domain wall picture and the chiral order should be closely related to each other.
Such an analysis of the connection between  chiral ordering and domain wall excitations may be an interesting problem.
Then, we should also  recall that the crossover between the interchain and intrachain processes of the excitations in the two-component TL liquid was  discussed for the integrable model, which has an explicit chirality term in its Hamiltonian\cite{frahm}.

Finally, we would like to note that the present method can work  effectively in the ferromagnetic region of the nearest-neighbor coupling $J_1<0$, where a similar chiral order can be expected.\cite{chubukov,honecker,kecke}
Although the zigzag chain is a simple model, we think that it still contains   important intrinsic  physics.


\acknowledgments

We would like to thank T. Hikihara for valuable discussions.
This work is supported by  Grants-in-Aid for Scientific Research from the Ministry of Education, Culture, Sports, Science and Technology of Japan (No. 18740230 and No. 20340096), and  by a Grant-in-Aid for Scientific Research on Priority Area ``High-field spin science in 100T''.
It is also partly supported by Priority Area ``Novel state of matter induced by frustration''.


\begin{thebibliography}{99}
\bibitem{dimer} The transition of the spin liquid and the dimer gapped state, and  the order disorder transition of the chirality are expected at the same $\alpha_c$ in Ref.[\cite{hikihara}].
In this paper, however, we basically concentrate on the chiral order.
\bibitem{current} 
Thus the definition of the spin current differs by twice from that in  Ref.[\cite{nersesyan}], where $\partial_x \Theta_+$ is associated with the sum of the currents of the double chains.
\bibitem{onset} We adopt the onset magnetization of the chiral order for a 192-spin system as a phase boundary. Thus our estimation of the chiral-ordered phase may be  slightly narrower than that in Ref. [\cite{mccolloch}].
\end{thebibliography}
\end{document}